%
% Author: Asish Satpathy
% University of Texas at Austin   
% BaBar Collaboration
% 
% Use this class for normal prd format
\documentclass[twocolumn,showpacs,aps,prl,superscriptaddress]{revtex4}
\usepackage{graphicx}
\usepackage{dcolumn}
\usepackage{amsmath}
\usepackage{epsfig}

\input pubboard/babarsym

\noindent \BABARPubNumber \\
\noindent \SLACPubNumber \\ 
%\noindent \LANLNumber \\

\begin{center}  \end{center}

\title{ {\large \bf \boldmath Search for the W-exchange decays {\boldmath $B^{0} \to D_{s}^{(*)-}  D_{s}^{(*)+}$}}}

%% author list as of 02-Aug-2005 (633 authors)
%% author list as of 02-Aug-2005 (633 authors)
%
\author{B.~Aubert}
\author{R.~Barate}
\author{D.~Boutigny}
\author{F.~Couderc}
\author{Y.~Karyotakis}
\author{J.~P.~Lees}
\author{V.~Poireau}
\author{V.~Tisserand}
\author{A.~Zghiche}
\affiliation{Laboratoire de Physique des Particules, F-74941 Annecy-le-Vieux, France }
\author{E.~Grauges}
\affiliation{IFAE, Universitat Autonoma de Barcelona, E-08193 Bellaterra, Barcelona, Spain }
% ...
\collaboration{The \babar\ Collaboration}
\noaffiliation

\begin{abstract}
We report a search for the decays \btodsds, \btodstds ~and \btodstdst ~in a
sample of 232 million $\Upsilon(4S)$ decays to \BBb ~pairs
collected with the \babar~detector at the PEP-II asymmetric-energy $e^+ e^-$ storage ring.
We find no significant signal and set upper bounds for the branching fractions:  
${\cal B}(B^{0} \to D_{s}^{-}  D_{s}^{+}) < 1.0 \times 10^{-4}, 
{\cal B}(B^{0} \to D_{s}^{*-}  D_{s}^{+}) < 1.3 \times 10^{-4}$ 
and ${\cal B}(B^{0} \to D_{s}^{*-}  D_{s}^{*+}) < 2.4 \times 10^{-4}$ at
90\% confidence level.
\end{abstract}

\pacs{ 13.25.Hw}% PACS, the Physics and Astronomy Classification Scheme.

\maketitle

In the Standard Model (SM), $B^{0} \to D_{s}^{(*)-}  D_{s}^{(*)+}$ decays 
are dominated by the W-exchange mechanism $\bar{b} d \to c \bar{c}$ as shown in
Figure 1, while the corresponding loop diagram is highly suppressed.
The decay rates of W-exchange or annihilation processes
are usually argued to be negligibly small due to the suppression from
helicity and/or form factors~\cite{fritzsch}; however this
assumption has not been well tested experimentally.

\begin{figure}[htb]
\setlength{\unitlength}{0.4mm}
\begin{picture}(70,200)(80,50)
%------------------------(1)
\put(70,240){\line(1,0){90}}
\put(70,190){\line(1,0){90}}
\put(62,236){$\bar{b}$}
\put(62,186){$d$}
\put(40,210.5){$B^0_d~~~~~~~~~~~W^{+}$}
\put(163,238){$\bar{c}$}
\put(163,187){$c$}
\put(163,225){$s$}
\put(163,200){$\bar{s}$}
\put(176,191){$D_{s}^{(*)+}$}
\put(176,230){$D_{s}^{(*)-}$}
\put(160,215){\oval(70,25)[l]}
\put(85,240){\vector(-1,0){2}}
\put(85,190){\vector(1,0){2}}
\put(145,240){\vector(-1,0){2}}
\put(145,190){\vector(1,0){2}}
\put(145,227.5){\vector(1,0){2}}
\put(145,202.5){\vector(-1,0){2}}
\multiput(100,235.6)(0,-4.1){12}{$>$}
\end{picture}
\vspace{-5.5cm}
\caption{\label{fig:dsds_dia} W-exchange decay diagram
for $B^0 \rightarrow D_{s}^{(*)-} D_{s}^{(*)+}$.}
\end{figure}
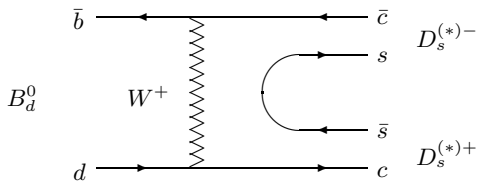
Recently, it has been pointed out that it is 
difficult to calculate these decay amplitudes using the factorization approach,
and a perturbative QCD (pQCD)~\cite{ying} model has been used to predict the
branching fraction for these decays. Prediction of branching fractions 
from an alternative model~\cite{eeg} gives
an estimate of non-factorizable contributions coming from chiral loops (CL) and
tree level amplitudes generated by soft gluon emission forming
a gluon condensate (GC) and it differs from pQCD approach 
by large amounts, as shown in Table~\ref{tab:theory-rev}.
\begin{table}[ht]
\caption{Summary of theoretical predictions of the branching fractions.}
\label{tab:theory-rev}
\begin{center} 
\begin{tabular}{ l c c }
\hline \hline
\B Decays & \multicolumn{2}{c}{Branching Fraction ($\times 10^{-5}$)} \\
%\cline{2-3}  
& ~~~pQCD~\cite{ying}~~~ & CL-GC~\cite{eeg} \\ \hline
&\\[-9pt]

$B^0 \rightarrow D_{s}^{-} D_{s}^{+}$ & $7.8 \pm ^{2.0}_{1.6}$ & $25.0$\\
$B^0 \rightarrow D_{s}^{*-} D_{s}^{+}$ & $6.0 \pm ^{1.6}_{1.1}$ &  $33.0$ \\
$B^0 \rightarrow D_{s}^{*-} D_{s}^{*+}$ & $8.5 \pm ^{2.0}_{1.8}$ & $54.0$ \\
\hline \hline 
\end{tabular} 
\end{center}
\end{table}

It has been estimated that a $C\!P$ asymmetry of the order of 10\% could arise
between $B^0 \rightarrow D_{s}^{-} D_{s}^{+}$ and its charge conjugate~\cite{blok}.
A measurement of the decay rates of $B^0 \to D_{s}^{(*)-} D_{s}^{(*)+}$
relative to those of $B^{0} \to D^{(*)-} D^{(*)+}$
will provide an estimate of the W-exchange contribution to the latter decay, a crucial
piece of information for extracting the CKM angle
$\gamma$ from $B^{0} \to D^{(*)-} D^{(*)+}$ 
and $B^0 \to D^{(*)-} D_{s}^{(*)+}$ decays~\cite{datta}.

Using 211 \invfb\ of data taken on the 
\FourS\ resonance with the \babar\ detector at the \pep2\ asymmetric $B$
factory, we report a search for 
$B^0 \rightarrow D_{s}^{-} D_{s}^{+}$,  $B^0 \rightarrow D_{s}^{*-} D_{s}^{+}$
and $B^0 \rightarrow D_{s}^{*-} D_{s}^{*+}$ decays~\cite{charge-conjugate}.
We use the $D_{s}^{*-}$ decays into $D_{s}^{-} \gamma $ and $D_{s}^{-}$ decays into $\phi \pi^{-}$, 
$K_{s}^{0} K^{-}$, and $K^{*0} K^{-}$. The
$\phi$, $K_{s}$ and $K^{*0}$ mesons are reconstructed in their decays to 
$K^{+}K^{-}, \pi^{+}\pi^{-}$ and \ksts, respectively. 

The \babar\ detector is described in detail
elsewhere~\cite{detector}. 
Tracking of charged particles is provided by a five-layer silicon
vertex tracker (SVT) and a 40-layer drift chamber (DCH). 
Discrimination between charged pions and kaons relies upon ionization energy loss (${\rm d}E/{\rm d}x$)
in the DCH and SVT, and upon Cherenkov photons detected in a ring-imaging
detector (DIRC). An electromagnetic
calorimeter (EMC), consisting of 6580 thallium-doped CsI crystals,
is used to identify electrons and photons. 
These detector sub-systems are mounted inside a 1.5-T solenoidal
superconducting magnet. Finally, the instrumented flux return of the 
magnet allows us to discriminate muons from other particles.
We use the GEANT4 Monte Carlo (MC)~\cite{geant4} program to simulate the
response of the detector, taking into account the varying
accelerator and detector conditions. 

Charged tracks used in the reconstruction of $\phi$, $K^{*}(892)^{0}$ and 
$D_{s}$ meson candidates must have a distance of closest approach to the 
interaction point of less than $1.5$~cm in the transverse plane and less than 
$10$ cm along the beam axis. 
All kaon candidates must pass 
particle identification (PID) criteria, based on a neural-network 
algorithm which uses measurements of ${\rm d}E/{\rm d}x$ in the DCH and the SVT, 
Cherenkov angles and the number of Cherenkov photons in the DIRC. 
No PID requirement is applied to the pion candidates. 
A $\phi$ candidate is composed of two identified kaons of opposite charge 
that are consistent with originating from a common vertex. We accept
$\phi$ candidates with invariant mass $1.000 < m_{K^{+}K^{-}} < 1.039$~\gev. 
$K_{s}^{0}$ candidates are composed of two oppositely-charged tracks
coming from a common vertex with an invariant mass 
$0.491 < m_{\pi^{+} \pi^{-}} < 0.505$~\gev. 
$K^{*}(892)^{0}$ candidates are reconstructed from 
two oppositely-charged tracks, where one track is identified as a kaon, with 
an invariant mass $ 0.842 < m_{K^{-}\pi^{+}} < 0.942 $~\gev.

We reconstruct $D_{s}^{-}$ mesons from 
decays to $\phi \pi^{-}$, $K_{s}^{0} K^{-}$, and $K^{*}(892)^0 K^{-}$
using tracks coming from a common vertex with a $\chi^2$ probability greater than 0.1\%.
The reconstructed mass of $D_{s}^{-}$ candidates is required to
be within 2.7 to 3.0 standard deviations of the nominal mass; 
a typical mass resolution of $D_{s}$ is about 5.1~\mev. 
The selected $D_{s}$ candidates are then 
kinematically fit with their masses constrained to the nominal value~\cite{PDG}. 
In the decays $D_{s}^{-} \to \phi \pi^{-}$ ($K^{*}(892)^0 K^{-}$),
the $\phi$ ($K^{*}(892)^0$) mesons are polarized longitudinally. Therefore
the cosine of the decay angle $\theta_{H}$ between the direction of
the $K^-$ from $\phi$ ($\pi^-$ from $K^{*}(892)^{0}$) and the $D_{s}^{-}$ 
direction in the $\phi$ ($K^{*}(892)^{0}$) rest frame is expected 
to follow $\cos^{2}\theta_{H}$ distribution.
Background events from random combinations are expected to be uniformly
distributed in $\cos\theta_{H}$.
We place a decay mode-dependent requirement on the minimum value of
$|\cos\theta_{H}|$, which varies from 0.3 to 0.5 and
rejects 13 to 24\% of the combinatorial background.

$D_{s}^{*-}$ candidates are formed by combining 
$D_{s}^{-}$ and $\gamma $ candidates with a mass
difference $\Delta M$ = $M_{D_{s}^{*-}} - M_{D_{s}^{-}} $ in the 
range of $0.125 < \Delta M <0.160 $ \gev. 
The photon energy measured in the EMC is required to 
be more than 100 \mev.

$B^{0}$ meson candidates are reconstructed by combining either
(i) two oppositely charged $D_{s}$ candidates, 
(ii) one $D_{s}^{*}$ candidate and an oppositely charged $D_{s}$ candidate 
or (iii) two oppositely charged $D_{s}^{*}$ candidates.
Finally, two quantities are used to discriminate between $B^{0}$-meson signal and 
background: the beam-energy-substituted mass  
$\mes = \sqrt{E^{*2}_{b} - ({\bf{p}^{*}_{B}})^{2}} $
and the energy difference $\Delta E = E^{*}_B - E^{*}_{b}$, 
where $E^{*}_{b}$ is the beam energy in the center of mass (CM) frame,
and ${\bf{p}^{*}_{B}}$ ($E^{*}_B$) is the CM momentum (energy) of the 
$B^{0}$-meson candidate.
For signal events \mes peaks at the $B^{0}$-meson mass with a typical resolution 
of 2.5 \mev, dominated by the uncertainty of the 
beam energy, and $\Delta E$ peaks near zero indicating that the
$B$ decay candidate has a total energy consistent with 
the beam energy in the CM frame. Depending on the particular $B^{0}$ decay mode,
the measured resolution for $\Delta E$ is $6.5-13.3$ \mev.

Multiple candidates are found in 3\% to 5\%
of the selected events in the three different $B^{0}$ decay modes.
The best candidate in each event is selected based on the smallest 
$\chi^2$ combination, where
\begin{eqnarray}\label{eq:chi2}
\chi^2\!\equiv \! \sum \left|\frac{m_{D_{s}^{\pm}}\!- \overline{m}_{D_{s}^{\pm}}}{\sigma_{m_{D_{s}^{\pm}}}}\right|^2\!\!+
\sum \left|\frac{\Delta M\!- \overline{\Delta M}}{\sigma_{\Delta M}}\right|^2\!\!,
\end{eqnarray}
and the sum is over $D_{s}^{(*)+}$ and $D_{s}^{(*)-}$ candidates participating in a particular 
$B^{0}$ decay.
The mean values ($\overline{m}_{D_{s}^{\pm}}$ and $\overline{\Delta M}$) are the nominal 
values given in Ref.~\cite{PDG} and the errors ($\sigma_{m_{D_{s}^{\pm}}}$, $\sigma_{\Delta M}$) 
are measured in a data control sample of $B^{0} \to D^{-} D_{s}^{(*)+}$ decays.

A small source of remaining background is $e^{+} e^{-} \to q \bar{q}$ production,
which is suppressed based on event topology.
We restrict the angle ($\theta_{T}$) between the thrust axis~\cite{thrust-axis} of the
$B^{0}$ meson candidate and the thrust axis of the rest of the particles in the event.
In the CM frame, \BBb~pairs are produced approximately at rest and
form a nearly uniform distribution in $|\cos\theta_{T}|$. In contrast, hadrons in $q \bar{q}$ events
are produced back-to-back in two jets, which results in a $|\cos\theta_{T}|$
distribution peaked at 1. Based on the background level of each mode, we require
the value of $|\cos\theta_{T}|$ to be less than a mode-dependent upper limit, 
which ranges from 0.83 and 0.9. We require $R_2 < 0.4$, 
where $R_2$ is the ratio of the second Fox-Wolfram moment to the zeroth 
moment~\cite{fox-wolfram}, both determined using charged tracks 
and unmatched neutral showers in the event. 

For different $B^{0}$ meson decays, a signal region is 
defined in a two dimensional scatter plane of \mes and
$\Delta E$ as shown in Table~\ref{tab:efficiency}. 
Optimization of the selection
is performed separately for each of the three $B^{0}$ decays~\cite{optimization} by
maximizing a figure of merit, $S^{2}/(S + B)$, where $S$ is the number of signal events
in the signal box as derived from the MC simulation and $B$ is the number of background events 
estimated from simulations of generic $B$-decays and $q \bar{q}$ continuum.
We use the same selection criteria for different $B^{0}$ decay modes if
the figure of merit differs by less than 10\%.

After the aforementioned selection, four possible background sources are considered. 
First, the amount of combinatorial background in the signal region is estimated
from the grand sideband region: $-0.25 < \Delta E <0.25$ \gev and $5.20 < m_{ES} < 5.27$ \gev.
The second source of backgrounds arises from
$B$ meson decays such as $B^{0} \to D^{(*)-} D_{s}^{(*)+}$ and 
$B^{-} \to D_{s}^{(*)-} D^{(*)0}$.
These background events have the same
\mes distribution as the signal, but their reconstructed energy is higher than the beam energy. 
Third, the cross-feed background that may arise among the six combinations of $D_{s} D_{s}$ modes
and the three reconstructed $B^{0}$ decay mode 
was studied with a large sample of signal MC and the corresponding
contributions were found to be small.
Finally, rare $B$ decays into the same final state particles, 
such as non-resonant $B^{0} \to D_{s}^{-} K^{0} K^{+}$, 
have the same \mes and $\Delta E$ distributions as the signal. This source of
background is estimated to be negligible.

Figure~\ref{fig:scatter} shows the distributions of candidates for
(i) $B^{0} \to D_{s}^{-} D_{s}^{+}$ (ii) $B^{0} \to D_{s}^{*-} D_{s}^{+}$ 
and (iii) $B^{0} \to D_{s}^{*-} D_{s}^{*+}$ decays
in the $\Delta E$ versus \mes plane after all selection criteria have been applied.
\begin{figure}[!htb]
\begin{center}
\includegraphics[width=8.0cm, height=5.2cm]{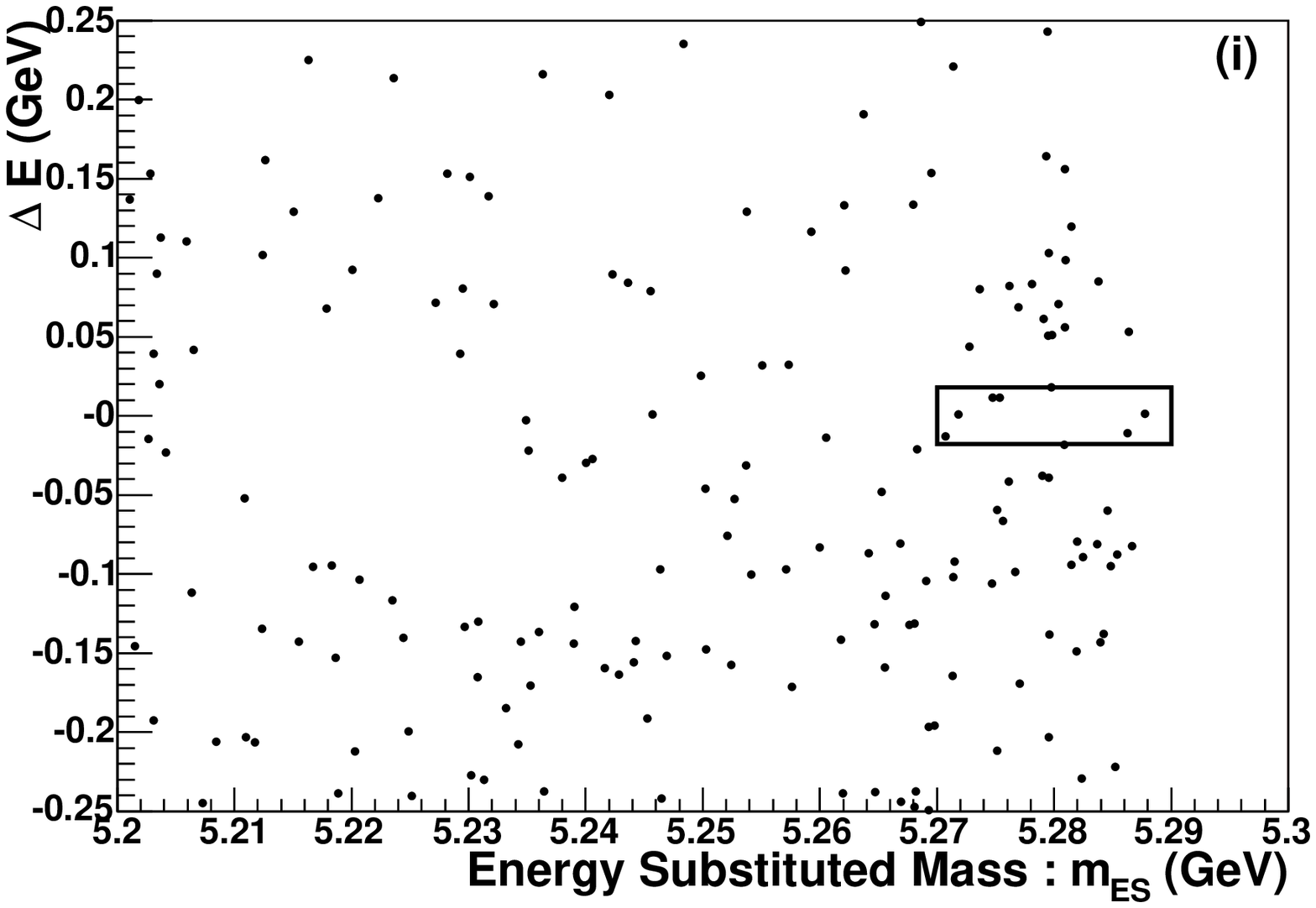}\\
\includegraphics[width=8.0cm, height=5.2cm]{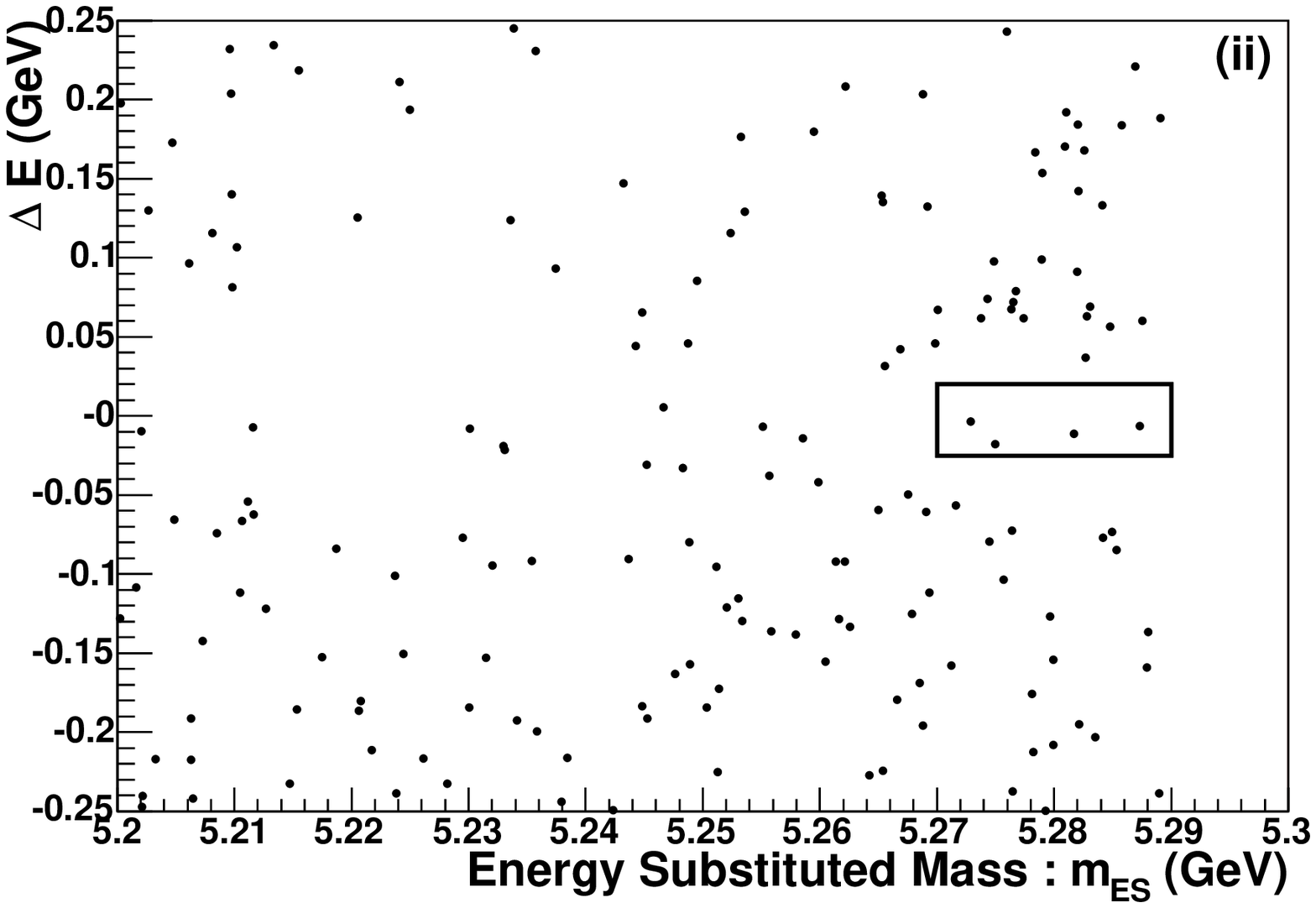}\\
\includegraphics[width=8.0cm, height=5.2cm]{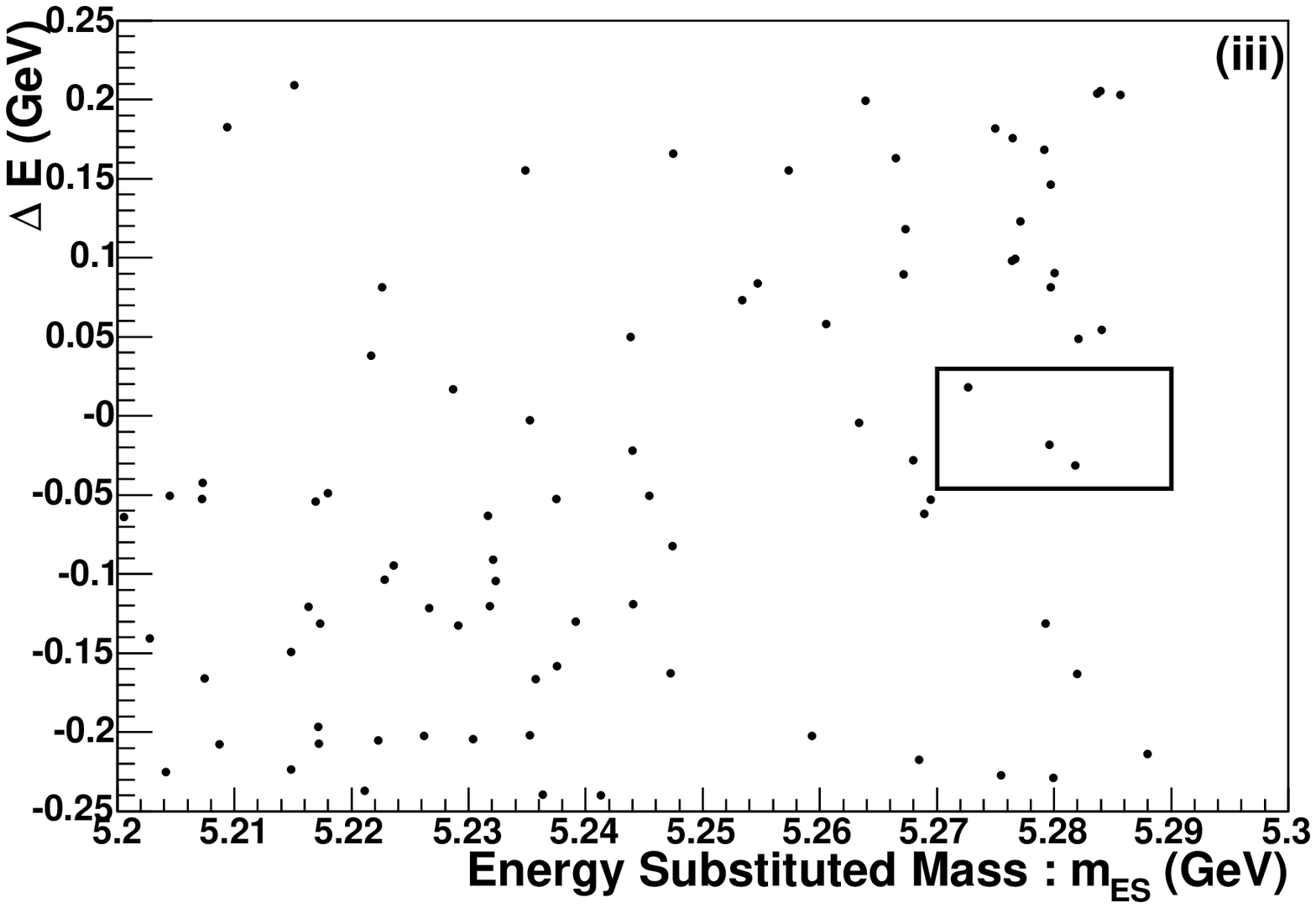}\\
\caption{Distributions of events in the $\Delta E$ versus \mes plane, for
(i) $B^{0} \to D_{s}^{-} D_{s}^{+}$ (ii) $B^{0} \to D_{s}^{*-} D_{s}^{+}$ 
and (iii) $B^{0} \to D_{s}^{*-} D_{s}^{*+}$ decays after all selection criteria are applied. 
The box in each plot is the signal region based on studies with MC simulation as 
described in the text.}
\label{fig:scatter}
\end{center}
\end{figure}
We find 6, 4 and 3 candidate events in the signal boxes 
that survived the selection criteria for the $B^{0} \to D_{s}^{-} D_{s}^{+}$, 
$B^{0} \to D_{s}^{*-} D_{s}^{+}$ and $B^{0} \to D_{s}^{*-} D_{s}^{*+}$ 
decay processes, respectively. The combinatorial background 
in the signal box ($N_{bkg}^{comb}$), is estimated from the number of events
in the grand sideband region of the data.
We compute the average number of background ($N_{bkg}^{avg}$) within the 
region $E_{2} < \Delta E <E_{1}$ \gev and $5.20 < m_{ES} < 5.27$ \gev 
from a fit to the 
$\Delta E$ distribution of the data events in the grand sideband (described well by 
a first order polynomial function $P(\Delta E)$) as: 
$N_{bkg}^{avg} = N_{GSB} \times \int^{E_{2}}_{E_{1}} P(\Delta E)/ \int^{0.25}_{-0.25} P(\Delta E) $,
where $E_{1}$ and $E_{2}$ are the $\Delta E$ energy bounds 
of the signal box as shown in Table~\ref{tab:efficiency} and $ N_{GSB}$ is the
total number of events in the grand sideband region.
The $m_{ES}$ projection of these background events is modeled with the
threshold function~\cite{argus}, 
\begin{eqnarray}\label{eq:threshold}
\frac{dN}{dx} = x\sqrt{1-{x}^2 / E_{b}^{*2}} {~\rm exp} [\xi(1-{x}^2/E_{b}^{*2})],
\end{eqnarray}
characterized by the shape parameter $\xi$, the 
endpoint parameter $E_{b}^{*}$ fixed at 5.289 \gev and $x = m_{ES}$.
$N_{bkg}^{comb}$ in the signal box is then estimated from $N_{bkg}^{avg}$ scaled 
by a factor: $\int^{5.29}_{5.27}\frac{dN}{dx}/ \int^{5.27}_{5.2} \frac{dN}{dx}$.
We vary $E_{b}^{*}$ by $\pm 2$ \mev 
to include its effect in the systematic uncertainties in $N_{bkg}^{comb}$. 
The measured uncertainties due to the choice of threshold parameter $\xi$,
endpoint parameter $E_{b}^{*}$, and parameter of the polynomial fit 
are combined in quadrature with the Poisson fluctuation of the number of events
in the grand sideband to obtain the total error on $N_{bkg}^{comb}$. 
This procedure does not account for any potential 
backgrounds that are enhanced in the 
signal region. The simulation indicates that only a 
small component of the background from the $B^{0} \rightarrow D_{s}^{(*)-}D^{(*)+}$ 
and $B^{0} \rightarrow D_{s}^{(*)-}D^{(*)0}$ decay 
exhibits a peaking \mes distribution. This component, $N_{bkg}^{peak}$, is 
extracted from a binned likelihood fit to the \mes distribution of simulated events
using a combination of the threshold function and a Gaussian.
The $\xi$ parameter in the threshold function is fixed to the
value we obtained from the fit to the data grand sideband. The mean
and width of the Gaussian component is fixed to the fit values obtained 
from $B^{0} \to D_{s}^{-} D^{+}$ decays in the data. Uncertainties in $N_{bkg}^{peak}$ arising
from the $D_{s}^{(*)-}D^{(*)+}$ and $D_{s}^{(*)-}D^{(*)0}$ branching fractions~\cite{PDG}
are added to its statistical error obtained from the fit. 
$N_{bkg}^{comb}$ and $N_{bkg}^{peak}$ are added 
to obtain the total estimated background, $N_{bkg}$, 
as quoted in Table~\ref{tab:efficiency}.
\begin{table}[ht]
\caption{ The number of signal candidates ($N_{cand}$), total estimated background 
($N_{bkg}$), efficiency from MC simulation times the branching fraction
($\epsilon_{i} \times {\cal B}$), and 90\% C.L. upper limit for 
$B^{0} \to D_{s}^{-} D_{s}^{+}$, $B^{0} \to D_{s}^{*-} D_{s}^{+}$ 
and $B^{0} \to D_{s}^{*-} D_{s}^{*+}$ decay modes.}
\label{tab:efficiency}
\begin{center}
\begin{tabular}{ l l l l }
\hline \hline
&\\[-9pt]
  &   $B^{0} \to D_{s}^{-} D_{s}^{+}$  & $B^{0} \to D_{s}^{*-} D_{s}^{+}$ &  $B^{0} \to D_{s}^{*-} D_{s}^{*+}$ \\
\hline
&\\[-9pt]
$\Delta E $(MeV)                              &\ -18.0--18.0     &\  -25.0--20.0     &\ -46.0--30.0   \\
\hline
\mes (GeV)                                    &\ 5.27--5.29     &\  5.27--5.29      &\ 5.27--5.29    \\
\hline
$\sum_{i} \epsilon {\cal B}$ &\ 3.51 $\times 10^{-4}$   &\  1.47 $\times 10^{-4}$   &\ 0.85  $\times 10^{-4}$  \\ 
\hline
$N_{cand}$                                 &\ 6             &\  4              &\ 3             \\
\hline
$N_{bkg}$                                     &\ $3.3\pm 1.0$  &\  $3.9 \pm 1.2$  &\ $2.3 \pm 0.9$ \\
\hline
U.L.                 &\ $< 1.0 \times 10^{-4}$      &\  $< 1.3 \times 10^{-4}$        &\ $< 2.4 \times 10^{-4} $   \\
&\\[-9pt]
\hline \hline
\end{tabular}
\end{center}
\end{table}

We consider the following sources of systematic uncertainty for the signal efficiencies. 
The particle reconstruction and identification efficiencies are obtained from simulation,
and cross-checked and corrected using large data control samples. This results in systematic
uncertainties of (1) 0.8\% per charged track; 
(2) 2.5\% per reconstructed $K^{0}_{s}$ candidate; (3) 2.5\% per identified charged kaon 
and (4) 1.8\% per reconstructed photon. 
The uncertainty on the number of \BBb ~ events is estimated to be 1.1\%.
Depending on the $B$ submodes, the error from the MC statistics is 2\% to 4.5\%.
The systematic errors are dominated by the 13.3\% 
relative uncertainty on ${\cal B}(D_{s}^{-} \to \phi \pi^{-})$~\cite{babar_new},
and 15.8\% and 9.8\% errors in ${\cal B}(D_{s}^{-} \to K_{s}^{0} K^{-})$ 
and ${\cal B}(D_{s}^{-} \to K^{*0} K^{-})$ relative to ${\cal B}(D_{s}^{-} \to \phi \pi^{-})$, 
respectively~\cite{PDG}. The uncertainty in modeling the simulation of the 
$\Delta E$, $|\cos\theta_{T}|$, $|\cos\theta_{H}|$ distributions
is evaluated using a ratio of the signal yield from 
$B^{0} \to D^{-} D_{s}^{(*)+}$ data control sample and generic \BBb ~MC.
Each selection requirement is varied and the resulting relative
change in the ratio is assigned as the systematic error.
The error due to vertexing is obtained 
by taking the difference in the ratio with and without the vertex requirement in the 
$D_{s}$ candidate selection. A summary of the systematic uncertainties in 
signal efficiency is given in Table~\ref{tab:systematics}.
\begin{table}[ht]
\caption{Summary of systematic uncertainties for signal efficiencies.}
\begin{center}
\begin{tabular}{l c c c}
\hline \hline
&\\[-9pt]
Systematics  & $D_{s}^{-} D_{s}^{+}$ (\%) & $D_{s}^{*-} D_{s}^{+}$ (\%)  & $D_{s}^{*-} D_{s}^{*+}$ (\%)  \\ 
\hline
%&&&\\[-9pt]
Tracking eff. & 4.3 & 4.3 & 4.3 \\
$K_{s}$ eff. & 2.7  & 2.7 & 2.7  \\
Kaon PID   & 9.2   & 9.2  & 9.2 \\
Photon eff. &  - & 1.8  & 3.6 \\ 
$B$ counting  & 1.1  & 1.1  & 1.1 \\
MC statistics & 2.0  & 3.5 &  4.5 \\ 
$D_{s}^{(*)}$ b.f. & 26.0  & 26.0 & 26.0 \\ 
Selection & 5.4 & 5.4 & 6.0 \\  \hline
Total   & 28.7  & 28.8 & 29.3 \\ 
&\\[-9pt]
\hline \hline
\end{tabular}
\end{center}
\label{tab:systematics}
\end{table}
Using the measured signal efficiency ($\sum_{i} \epsilon_{i} {\cal B}_{i}$), 
211 $\fb^{-1}$ on-resonance data corresponding to 
$N_{\B\Bbar}$ = (231.8 $\pm$ 2.6) $\times 10^{6}$, the background estimation along
with the uncertainties and the observed candidate events in the signal region $N_{cand}$, 
we determine the 90\% confidence-level (C.L.) upper limit using the procedure given 
in~\cite{RogerBarlow}. The systematic
uncertainties are included following the prescription in 
Ref.~\cite{CousinHighLand}. In all branching fraction calculations
we assume equal production of \BzBzb 
and $B^{+}B^{-}$ pairs at the $\Upsilon(4S)$.

The search for $B^{0} \to D_{s}^{-} D_{s}^{+}$, $B^{0} \to D_{s}^{*-} D_{s}^{+}$ 
and $B^{0} \to D_{s}^{*-} D_{s}^{*+}$ decays yields 
the 90\% C.L. upper limits (Table~\ref{tab:efficiency}):
\begin{eqnarray*}
{\cal B}(B^{0} \to D_{s}^{-} D_{s}^{+}) &<& 1.0 \times 10^{-4}, \\
{\cal B}(B^{0} \to D_{s}^{*-} D_{s}^{+}) &<& 1.3 \times 10^{-4}, \\
{\cal B}(B^{0} \to D_{s}^{*-} D_{s}^{*+}) &<& 2.4 \times 10^{-4}.
\end{eqnarray*}
In conclusion, we have performed a measurement of the decay rates for
$B^{0} \to D_{s}^{-} D_{s}^{+}$, $B^{0} \to D_{s}^{*-} D_{s}^{+}$ 
and $B^{0} \to D_{s}^{*-} D_{s}^{*+}$ processes with a  
sensitivity needed to test the SM prediction \cite{BelleDsDs}. Our
upper limits disfavor the branching fraction predictions in Ref.~\cite{eeg} for 
all three $B^{0}$ decays and accommodate the predictions of the pQCD calculation~\cite{ying}
for all three $B^{0}$ decay modes. The possible existence of a significant W-exchange component
in $B^{0} \to D^{-} {D^{+}}$~\cite{Gobinda} decays is not confirmed 
in this analysis.

We are grateful for the 
extraordinary contributions of our \pep2\ colleagues in
achieving the excellent luminosity and machine conditions
that have made this work possible.
The success of this project also relies critically on the 
expertise and dedication of the computing organizations that 
support \babar.
The collaborating institutions wish to thank 
SLAC for its support and the kind hospitality extended to them. 
This work is supported by the
US Department of Energy
and National Science Foundation, the
Natural Sciences and Engineering Research Council (Canada),
Institute of High Energy Physics (China), the
Commissariat \`a l'Energie Atomique and
Institut National de Physique Nucl\'eaire et de Physique des Particules
(France), the
Bundesministerium f\"ur Bildung und Forschung and
Deutsche Forschungsgemeinschaft
(Germany), the
Istituto Nazionale di Fisica Nucleare (Italy),
the Foundation for Fundamental Research on Matter (The Netherlands),
the Research Council of Norway, the
Ministry of Science and Technology of the Russian Federation, and the
Particle Physics and Astronomy Research Council (United Kingdom). 
Individuals have received support from 
CONACyT (Mexico),
the A. P. Sloan Foundation, 
the Research Corporation,
and the Alexander von Humboldt Foundation.

%
% --- bibliography
%   


\begin{thebibliography}{99}
\bibitem{fritzsch} H. Fritzsch and P. Minkowski, \jplb {90}, 455 (1980);
  D. Fakirov and B. Stech, Nucl.~Instr.~Methods~Phys.~Res., Sect.~B {\bf 133}, 315 (1978).
\bibitem{ying} Ying Li {\it et al.} J.~Phys. {\bf G \bf 31}, 273 (2005). 
%A general concept on perturbavite QCD approach 
%can be found at: Yong-Yeon Keum {\it et al.} \prd {\bf 63}, 054008 (2001).
\bibitem{eeg} J. O. Eeg {\it et al.} \epjc {\bf 42}, 29 (2005).
\bibitem{blok} B. Blok {\it et al.} \jprl {\bf 78}, 3999 (1997).
\bibitem{CKM} The \babar\ Physics Book, P. Harrison and H. Quinn, 1998, SLAC report 504.
\bibitem{datta} A. Datta, D. London \jplb {\bf 584}, 81 (2004).
\bibitem{charge-conjugate} Inclusion of charge conjugate modes is implied throughout this paper.
\bibitem{detector} \babar\ Collaboration, B.~Aubert {\em et al.}, Nucl.~Instr.~Methods~Phys.~Res., Sect.~A {\bf 479}, 1 (2002).
\bibitem{geant4} GEANT4 Collaboration, S.~Agostinelli {\em et al.}, Nucl.~Instr.~Methods~Phys.~Res., Sect.~A {\bf 506}, 250 (2003).
\bibitem{PDG} Particle Data Group, S.~Eidelman {\it et al.}, \jplb {\bf 592}, 1 (2004).
\bibitem{thrust-axis} E. Farhi, \jprl{\bf 39}, 1587 (1977).
\bibitem{fox-wolfram} G. C. Fox, S. Wolfram, \jprl {\bf 41}, 1581 (1978).
\bibitem{optimization}pQCD predictions for the branching fractions are assumed for the selection optimization.
\bibitem{argus} ARGUS Collaboration, H.~Albrecht {\em et al.}, \zp {\bf C48}, 543 (1990).
\bibitem{babar_new} \babar\ collaboration, B. Aubert {\it et al.}, \prd {\bf 71}, 091104(R) (2005).
\bibitem{RogerBarlow} R. Barlow, Comput. Phys. Commun. {\bf 149}, 97 (2002).
\bibitem{CousinHighLand} R. D. Cousins and V. L. Highland, Nucl.~Instr.~Methods~Phys.~Res., Sect.~A {\bf 320}, 331 (1992).
\bibitem{BelleDsDs}While this paper was being written, the Belle Collaboration released an upper limit on the 
decay rate for $B^{0} \to D_{s}^{-} D_{s}^{+}$ consistent with our result: K. Abe {\it et. al.} hep-ex/0508040. 
\bibitem{Gobinda}Belle Collaboration, G. Majumder {\it et. al.} \jprl {\bf 95}, 041803 (2005).
\end{thebibliography}
\end{document}